\documentstyle[11pt]{article}
\newcommand{\be}{\begin{eqnarray}}

\newcommand{\ee}{\end{eqnarray}}

\title{
        \begin{flushright}
        {\normalsize
        NBI--97--26\\
        June 1997 \\}
        \end{flushright}
\bf     On colliding ultrarelativistic nuclei on a transverse lattice
       }
\author{Alex Krasnitz and Raju Venugopalan\\ 
        {\small\it Niels Bohr Institute,
        Blegdamsvej 17,
        Copenhagen, Denmark, DK--2100 } \\          
       }

\date{}

\begin{document}
\setcounter{page}{0}
\maketitle
\thispagestyle{empty}
\begin{center}
{\bf Abstract}\\
\end{center}

\noindent 
We argue that the classical evolution of small x modes in the collision of
two ultrarelativistic nuclei is described on a transverse lattice by the 
Kogut--Susskind Hamiltonian in
2+1-dimensions coupled to an adjoint scalar field. The initial conditions
for the evolution are provided by the non--Abelian Weizs\"acker--Williams
fields which constitute the classical parton distributions in each of the
nuclei. We outline how lattice techniques developed for real time simulations 
of field theories in thermal equilibrium can be used to study 
non--perturbatively, 
thermalization and classical gluon radiation in ultrarelativistic nuclear
collisions.

\vfill \eject

\section{Introduction}
\vspace*{0.3cm}

It is of considerable theoretical and experimental interest to understand 
the collisions of nuclei at ultrarelativistic energies and the putative
evolution of the hot and dense matter created in these collisions into 
a thermalized, deconfined state of matter called a quark gluon plasma.
The theoretical challenge is to understand the dynamics of the formation of 
this matter
from QCD, and its properties, while the experimental challenge is to detect 
evidence that such a plasma was indeed formed~\cite{QM96}.

The space--time evolution of the nuclei after the collision and the 
magnitudes and
relevance of various proposed signatures of this hot and dense matter 
depend sensitively on the initial conditions for the evolution, namely,
the parton distributions in each of the nuclei {\it prior} to the 
collision. In the conventional perturbative QCD approach to
the problem, observables from the collision may be computed by convolving
the parton distributions of each nucleus, determined from deep 
inelastic scattering experiments, with the elementary parton--parton 
scattering cross sections. The cross sections thereby obtained are often
incorporated either in a multiple scattering formalism~\cite{Wang} or in a 
classical cascade approach to obtain the space--time evolution~\cite{Geiger}.
While the above approach provides a reasonable description of large transverse 
momentum processes at large $x$, it is not sufficient as we go to small $x$ 
or alternatively, towards central rapidities~\cite{comment}. This is because 
at small $x$ 
partons in one nucleus may ``see'' more than one parton in the direction
of the incoming nucleus resulting in a breakdown of the above described 
convolution of distributions. What is needed therefore to describe the
collision of the ``wee'' nuclear partons is, roughly put, products of 
amplitudes as opposed to products of probabilities.

A model describing the small $x$ modes in large nuclei was formulated 
by McLerran and Venugopalan~\cite{RajLar}. The model contains one dimensionful
parameter, $\chi(y,Q^2)$, which is the total color charge squared per unit 
area 
integrated from the rapidity $y$ of interest to the beam rapidity. Since it is
the only scale in the problem, the coupling constant runs as a function of
this scale. One therefore has weak coupling in the limits where the color
charge $\chi$ is large; either for $A>>1$ or $s\rightarrow \infty$. 
It was argued there that the 
classical background fields in this model are non--Abelian 
Weizs\"acker--Williams fields. Exact analytical expressions for these fields
have been obtained recently~\cite{AlexLar,Kovchegov}. Further, it has been
shown explicitly that $\chi$ obeys renormalization group equations in $y$ 
and $Q^2$ which 
reduce to the well known DGLAP and BFKL equations~\cite{BFDGL} in the 
appropriate limits~\cite{AlexLar,Kovner}.

The model was applied to the problem of nuclear collisions
by Kovner, McLerran and Weigert, who formulated the problem as the collision
of Weizs\"acker--Williams fields~\cite{KLW}. The classical background fields
after the collision then correspond to solutions of the Yang--Mills equations 
in the presence of static, random sources of color charge on the light cone. 
The classical background field
for the two nuclei after the collision was found and perturbative solutions
obtained for modes with transverse momenta $k_t>>\alpha_S\sqrt{\chi}$. 
After averaging over the Gaussian random light cone sources, the energy 
and number distributions of physical gluons were computed. Further, the 
classical gluon radiation from these perturbative modes was studied by these 
authors and later in greater detail by several 
others~\cite{gyulassy,DirkYuri,SerBerDir}. While the perturbative approach is
very relevant and useful, it is still essential to consider the full non--
perturbative approach for the following reasons. Firstly, the classical 
gluon radiation computed perturbatively is infrared singular and has to
be cut-off at some scale. It was argued in Ref.~\cite{KLW,DirkYuri} that a 
natural scale where the distributions are cut-off is given by $k_t\sim 
\alpha_S \sqrt{\chi}$. However, since quantitative differences can be large, it
is important to perform a full calculation. Secondly, the non--perturbative
approach is crucial to study the possible thermalization of the system and
the relevant time scales for thermalization. This in turn has several ramifications for computations of various signatures of the quark gluon plasma. For 
instance, if thermalization does occur, then as proposed by Bjorken~\cite{Bj} 
hydrodynamic evolution of the system is reasonable. In that event, our 
approach would provide the initial temperature and velocity profiles 
necessary for such an evolution (see ~\cite{josef} and references therein).

In this paper, we outline how one may perform real--time simulations of
the full, non--perturbative evolution of classical non--Abelian 
Weizs\"acker--Williams fields. Such a simulation is possible since the fields 
are  classical. Similar real time simulations of classical fields have been 
performed in the context of sphaleron induced baryon number violation~
\cite{Krasnitz} and chirality violating transitions in hot gauge theories
~\cite{Moore}. In brief, the idea is as follows. We write down the lattice
Hamiltonian which describes the evolution of these classical gauge fields. 
It turns out to be the Kogut--Susskind Hamiltonian in 2+1--dimensions 
coupled to an adjoint scalar field. The lattice equations of motion for the 
fields are thereby determined straightforwardly. The initial conditions for
the evolution are provided by the Weizs\"acker--Williams fields for the
nuclei before the collisions. Interestingly, the dependence on the static
light cone sources does not enter through the Hamiltonian but instead from
the initial conditions. Also, to reiterate, our results have to be averaged
over by the above mentioned Gaussian measure.

The paper is organized as follows. In the following section we briefly
discuss the problem of initial conditions as formulated by Kovner, McLerran
and Weigert and their perturbative solution. In section 3, we derive the
expression for the lattice Hamiltonian. The lattice equations of motion and
initial conditions are discussed in section 4. We conclude with a discussion
of observables that can be computed on the lattice and comment on checks that
can be performed on such computations.

\section{The classical background field of two nuclei on the light cone}
\vspace*{0.3cm}

In the work of McLerran and Venugopalan~\cite{RajLar}, the classical gluon 
field at small $x$
for a nucleus in the infinite frame is obtained by solving the Yang--Mills 
equations in the presence of a static source of color charge 
$\rho^a(r_t,\eta)$ on the 
light cone. Exact solutions for the classical field as functions of $\rho^a 
(r_t,\eta)$ were found by Jalilian--
Marian et al.~\cite{AlexLar} and independently by Kovchegov~\cite{Kovchegov}.
Distribution functions are computed by averaging over products of the 
classical fields over a Gaussian measure in $\rho$ with the variance 
$\mu^2(\eta,Q^2)$. Here $\mu^2$ is the color charge squared per unit area per
unit rapidity resolved at a scale $Q^2$ by an external probe. It is related
to $\chi$ by the expression $\chi(\eta,Q^2)=\int_\eta^\infty \mu^2(
\eta^\prime,Q^2)$.

The above picture of gluon fields in a nucleus at small $x$ was extended
to describe nuclear collisions by Kovner, McLerran and Weigert~\cite{KLW}. 
We shall review and discuss their paper below. 
The classical background field of two nuclei is described by the Yang--Mills 
equations in the presence of two light cone sources--one on each light cone.
We have then
\be
D_\mu F^{\mu\nu} = J^\nu \, ,
\label{yangmill}
\ee
where 
\be
J^{\nu,a}(r_t) = \delta^{\nu +}g\rho_{+}^a (r_t)\delta(x^-) + \delta^{\nu -}
g\rho_{-}^a (r_t) \delta(x^+) \, .
\label{sources}
\ee
Gluon distributions are simply related to the Fourier transform $A_i^a (k_t)$ 
of the solution to the above equation by $<A_i^a(k_t) A_i^a(k_t)>_\rho$. The
averaging over the classical charge distributions is defined by
\be
\langle O\rangle_\rho &=& \int d\rho_{+}d\rho_{-}\, O(\rho_+,\rho_-) \nonumber \\
&\times& \exp\left( -\int d^2 r_t {{\rm Tr}\left[\rho_+^2(r_t)+\rho_-^2(r_t)
\right]
\over {2\mu^2}}\right) \, .
\ee
The averaging over the color charge distributions is performed independently 
for each nucleus with equal Gaussian weight $\mu^2$.

The observant reader will notice that we have omitted the rapidity dependence
of the the charge distributions in the equations immediately above. We will
justify this omission in our discussion of the lattice Hamiltonian. We note
that the rapidity dependence of the charge distribution is also absent in 
Ref.~\cite{KLW}. 

Before the nuclei collide ($t<0$), a solution of the equations of motion is
\be
A^{\pm}&=&0 \, , \nonumber \\
A^i&=& \theta(x^-)\theta(-x^+)\alpha_1^i(r_t)+\theta(x^+)\theta(-x^-)
\alpha_2(r_t) \, ,
\ee
where~\cite{gyulassy}
\be
\alpha_{1,2}^i(r_t) = {1\over {ig}}\left(Pe^{-ig\int_{\pm \eta_{proj}}^0 
d\eta^\prime
{1\over {\nabla_{\perp}^2}}\rho_{\pm}(\eta^\prime,r_t)}\right)^{\dagger} 
\nabla^i\left(Pe^{-ig\int_{\pm \eta_{proj}}^0 d\eta^\prime {1\over 
{\nabla_{\perp}^2}}\rho_{\pm}(\eta^\prime,r_t)}\right) \, .
\ee
Above, $\eta=\eta_{proj}-\log(x^-/x_{proj}^-)$ is the rapidity of the nucleus 
moving along the positive light cone with the gluon field $\alpha_1^i$ and
$\eta=-\eta_{proj}+\log(x_{proj}^+/x^+)$ is the rapidity of the nucleus moving
along the negative light cone with the gluon field $\alpha_2^i$. It is 
expected that at central rapidities (or $x<<1$) the source density varies
slowly as a function of rapidity and $\alpha^i\equiv \alpha^i(r_t)$.
The above expression suggests that for $t<0$ the solution is simply the
sum of two disconnected pure gauges.

For $t>0$ the solution is no longer pure gauge. Working in the Schwinger 
gauge
\be
x^+ A^- + x^- A^+ =0 \, ,
\label{schwinger}
\ee
the  authors of Ref.~\cite{KLW} found that with the ansatz
\be
A^{\pm}&=&\pm x^{\pm}\alpha(\tau,r_t)\, , \nonumber \\
A^i&=&\alpha_\perp^i(\tau,r_t) \, ,
\label{ansatz}
\ee
where $\tau=\sqrt{2x^+ x^-}$, Eq.~\ref{yangmill} could be written in
the simpler form
\newpage
\be
{1\over \tau^3}\partial_\tau \tau^3 \partial_\tau \alpha + [D_i,\left[D^i,
\alpha\right]]
&=&0 \, , \nonumber \\
{1\over \tau}[D_i,\partial_\tau \alpha_\perp^i] + ig\tau[\alpha,\partial_
\tau \alpha] &=&0\, ,\nonumber \\
{1\over \tau}\partial_\tau \tau\partial_\tau \alpha_\perp^i
-ig\tau^2[\alpha,\left[D^i,\alpha\right]]-[D^j,F^{ji}]&=&0 \, . 
\label{yangmill2}
\ee 
The initial conditions for the fields $\alpha(\tau,r_t)$ and $\alpha_\perp^i$ 
are given in terms of the fields for each of the nuclei at $t<0$. We have
\be
\alpha_\perp^i|_{\tau=0}&=& \alpha_1^i+\alpha_2^i \, , \nonumber \\
\alpha|_{\tau=0}&=&{ig\over 2} [\alpha_1^i,\alpha_2^i] \, .
\label{initial}
\ee
Further, since the equations are very singular at $\tau=0$, the only 
condition on the derivatives of the fields that would lead to regular
solutions are $\partial_\tau \alpha|_{\tau=0},\partial_\tau \alpha_\perp^i
|_{\tau=0} =0$. 

In Ref.~\cite{KLW}, solutions were found in the perturbative limit 
by expanding 
the initial conditions and the fields in powers $\rho$ or equivalently,
in powers of $\alpha_S\mu/k_t$. Performing a gauge transformation of the
above fields (such that the new fields satisify the Coulomb gauge condition),
at late times $\tau>>\alpha_S\mu$, the new fields can 
be expanded as Fourier series with coefficients $a_{1,2}^b(k_t)$ 
and their complex conjugates. The multiplicity distribution of classical
gluons can then be written as
\be
{dN\over {dy d^2k_t}} = {1\over {(2\pi)^3}}\sum_{i,b} |a_i^b(k_t)|^2 \, .
\ee
Detailed expressions for classical gluon radiation in the perturbative 
limit were obtained in Refs.~\cite{KLW,gyulassy,DirkYuri}. However, all these
expressions are infrared singular and have to be regulated by an infrared
cutoff. One advantage of solving the Yang--Mills equations to all orders
in $\alpha_S\mu/k_t$ is that it will likely provide a self--consistent, 
infrared safe result for the multiplicity of classical gluon radiation 
in ultrarelativistic nuclear collisions.

\section{Derivation of lattice Hamiltonian}
\vspace*{0.3cm}

While the Yang--Mills equations discussed above can be solved perturbatively
in the limit $\alpha_S k_t << \mu$, it is unlikely that a simple analytical 
solution exists for Eq.~\ref{yangmill} in general. The classical solutions
have to be determined numerically for $t>0$. The straightforward procedure 
would be to discretize Eq.~\ref{yangmill} but it will be more convenient 
for our purposes to construct the lattice Hamiltonian and obtain the lattice
equations of motion from Hamilton's equations.

We start from the QCD action (without dynamical quarks) which is given by
\be
S_{QCD} = \int d^4 x \sqrt{-g}\left\{{1\over 4} g^{\mu\lambda}g^{\nu\sigma}
F_{\mu\nu}F_{\lambda\sigma}-j^\mu A_\mu \right\} \, ,
\ee
where $g=det|g_{\mu\nu}|$. In the forward light cone ($t>0$) it is 
convenient to work with the $\tau,\eta,\vec{r_t}$ co--ordinates where 
$\tau=\sqrt{
2 x^+ x^-}$ is the proper time, $\eta={1\over 2}\log(x^+/x^-)$ is the 
space--time rapidity and $\vec{r_t}=(x,y)$ are the two transverse Euclidean 
co--ordinates. In these co--ordinates, the metric is diagonal with 
$g^{\tau\tau}=-g^{xx}=-g^{yy}=1$ and $g^{\eta\eta}=-1/\tau^2$.

After a little algebra, the Hamiltonian can be written as~\cite{Sasha}
\be
H =\int d\eta d\vec{r_t} \tau \left\{{1\over 2} p^{\eta}p^{\eta}
+{1\over {2\tau^2}}p^r p^r + {1\over{2\tau^2}} F_{\eta r}F_{\eta r}
+{1\over 4}F_{xy}F_{xy} + j^\eta A_\eta + j^r A_r\right\} \, .
\label{hamilton}
\ee
Here we have adopted the gauge condition of Eq.~\ref{schwinger}, which is 
equivalent to requiring $A^\tau =0$. 
Also, $p^\eta={1\over \tau}\partial_\tau A_\eta$ and $p^r=\tau \partial_\tau 
A_r$ are the conjugate momenta.

Consider the field strength $F_{\eta r}$ in the above Hamiltonian. If we 
assume approximate boost invariance, or
\be 
A_r (\tau,\eta,\vec{r_t})\approx A_r(\tau,\vec{r_t}); \ \  
A_{\eta}(\tau,\eta,\vec{r_t})\approx \Phi(\tau,\vec{r_t}),
\ee
we obtain  
\be
F_{\eta r}^a = -D_r \Phi^a \, ,
\label{fdstrgth}
\ee
where $D_r =\partial_r -ig A_r$ is the covariant derivative. Further, if we 
express $j^{\eta,r}$ in terms of the $j^{\pm}$ defined in 
Eq.~\ref{sources} we obtain the result that $j^{\eta,r}=0$ for $\tau>0$.
Finally, since 
\be
\Phi = \tau^2 \alpha(\tau,\vec{r_t})\, ; \, A_r = \alpha_{\perp}^r 
(\tau, \vec{r_t})\, ,
\label{adef}\ee
we can perform the integration over the space--time rapidity to re--write
the Hamiltonian in Eq.~\ref{hamilton} as 
\be
H = \int d\vec{r_t} \tau \eta \left\{ {1\over 2}\left({\partial 
\alpha_{\perp}^r\over
{\partial\tau}}\right)^2 + {1\over 4}F_{xy}^a F_{xy}^a + {1\over {2\tau^2}}
\left(D_\beta \left[\tau^2 \alpha\right]\right)^2\right\} \, .
\label{twodh}
\ee
Here the index $\beta=(\tau,\vec{r_t})$. We have thus succeeded in
expressing the Hamiltonian in Eq.~\ref{hamilton} as the Yang--Mills Hamiltonian
in 2+1--dimensions coupled to an adjoint scalar. The discrete version
of the above Hamiltonian is well known and is the Kogut--Susskind 
Hamiltonian~\cite{KS} in 2+1--dimensions coupled to an adjoint scalar field. 
We shall now
restrict the discussion to an SU(2) Yang--Mills theory, in order to 
keep
notation simple. Generalization to an arbitrary SU(N) gauge group is
straightforward. The Kogut-Susskind analogue of Eq.~\ref{twodh} is
\be
H_L&=& {1\over{2\tau}}\sum_{l\equiv (j,\hat{n})} 
E_l^{a} E_l^{a} + \tau\sum_{\Box} \left(1-
{1\over 2}{\rm Tr} U_{\Box}\right)  \, ,\nonumber \\
&+& {1\over{4\tau}}\sum_{j,\hat{n}}{\rm Tr}\,
\left(\Phi_j-U_{j,\hat{n}}\Phi_{j+\hat{n}}
U_{j,\hat{n}}^\dagger\right)^2 +{\tau\over 4}\sum_j {\rm Tr}\,p_j^2,
\label{hl}\ee
where $E_l$ are generators of right covariant derivatives on the group
and $U_{j,\hat{n}}$ is a component of the usual SU(2) matrices corresponding
to a link from the site $j$ in the direction $\hat{n}$. The first two terms
correspond to the contributions to the Hamiltonian from the chromoelectric and
chromomagnetic field strengths respectively. In the last equation
$\Phi\equiv \Phi^a\sigma^a$ is the adjoint scalar field with its conjugate
momentum $p\equiv p^a\sigma^a$.

Finally, we should comment on a key assumption in the above derivation, 
namely, the boost invariance of the fields.
This invariance results in Eq.~\ref{twodh} thereby allowing us to restrict 
ourselves to a transverse lattice alone. To clarify the issue we are compelled
to make a few historical remarks. As we mentioned earlier, the authors 
of Ref.~\cite{KLW} found a solution which was explicitly boost invariant. 
However, this result was a consequence of the original assumption of McLerran 
and Venugopalan that the 
color charge density factorizes, $\rho^a(r_t,\eta)\rightarrow \rho^a(r_t) 
\delta(x^-)$. It was noticed in Ref.~\cite{Rajs}, this factorized 
form for the charge density results in infrared singular correlation functions 
which diverge as the square of the lattice size.
This problem was resolved in Ref.~\cite{AlexLar} where the authors realized 
that a rapidity dependent charge density $\rho^a(r_t,\eta)$ would give 
infrared safe solutions. This might be interpreted as implying that the 
boost invariance assumption of Ref.~\cite{KLW} should be given up as well.

Fortunately, this is not necessary. In principle, the rapidity dependence 
of the color charge density can 
be arbitrarily weak since that is sufficient to obtain infrared safe 
correlation functions. In Ref.~\cite{RajLar2}, an explicit model was 
constructed for the color charge distribution in the fragmentation region. 
It was shown there that for $\eta<\eta_{proj}$ the color charge distribution 
had a very weak dependence on $\eta$. We should note too that it was shown 
recently by Gyulassy and McLerran~\cite{gyulassy} that the initial conditions 
in Eq.~\ref{initial} are unaffected by the smearing in rapidity.

\section{Lattice equations of motion and initial conditions}
\vspace*{0.3cm}

Lattice equations of motion follow directly from $H_L$ of Eq.~\ref{hl}. 
For any
dynamical variable $v$ with no explicit time dependence ${\dot v}=\{H_L,v\}$,
where ${\dot v}$ is the derivative with respect to $\tau$, and $\{\}$ denote
Poisson brackets. We take $E_l$, $U_l$, $p_j$, and $A_j$ as independent 
dynamical variables, whose only nonvanishing Poisson brackets are
$$\{p_i^a,\Phi_j^b\}=\delta_{ij}\delta_{ab}; \ \ 
\{E_l^a,U_m\}=-i\delta_{lm}U_l\sigma^a; \ \
\{E_l^a,E_m^b\}=2\delta_{lm}\epsilon_{abc}E_l^c$$
(no summing of repeated indices). The equations of motion are consistent with
a set of local constraints (Gauss' laws). These are
\begin{equation}
C^a_j\equiv \sum_{\hat{n}}
\left[{1\over 2}E_{j,{\hat{n}}}^b{\rm Tr}\left(\sigma^aU_{j,{\hat{n}}}\sigma^b
U_{j,{\hat{n}}}^\dagger\right)
-E_{j-{\hat{n}},{\hat{n}}}^a\right]
-2\epsilon_{abc}p^b_jA^c_j=0.
\label{calpha}\end{equation}

The initial conditions for the fields and momenta in 
Eq.~\ref{initial} can be discretized as follows. To every lattice site $j$ we
assign two SU(2) matrices, $V_{1,j}$ and $V_{2,j}$.
Each of these two defines a lattice gauge
field configuration (corresponding to the two pure gauges discussed in 
section 2) whose link variables are 
$${\cal U}^q_{j,{\hat n}}\equiv V_{q,j}V^\dagger_{q,j+{\hat n}},$$
where $q=1,2$ labels the two nuclei. An initial condition for the link 
matrices, having the correct formal continuum limit, is then 
\be
U_{j,\hat{n}}|_{\tau=0}={{\sum_q {\cal U}^q_{j,{\hat n}}-I}\over
{\left({1\over 2}{\rm Tr}(\sum_q {\cal U}^q_{j,{\hat n}}-I)
(\sum_q {\cal U}^{q\dagger}_{j,{\hat n}}-I)\right)^{1/2}}},
\ee
where $I$ is the identity matrix.
Introducing $\alpha_q$ through $\exp(-ia\alpha_q)={\cal U}^q$, one can
easily verify that for smooth initial field configurations
$$1-{1\over 2}U_\Box\rightarrow {a^4\over 2}B^\nu B^\nu$$
in the limit of vanishing lattice spacing $a$. Here $B^\nu$ is the $\nu$th
color component of the SU(2) magnetic field corresponding to the gauge
potential $\alpha_1+\alpha_2$. Thus the initial plaquette term of 
Eq.~\ref{hl} has the correct formal continuum limit.

As we have already discussed, regularity requires the vanishing of the initial 
transverse color electric fields. If follows from Eq.~\ref{adef} that the 
adjoint scalar $A$ must also vanish initially. On the other hand, values of 
$p$, the
conjugate momentum of $A$, need not be zero at $\tau=0$. In the continuum
$$p^\eta|_{\tau=0}=2\alpha.$$
This initial condition can be written on the lattice as follows:
$$p_j|_{\tau=0}=-i\sum_{\hat n}\left(
[{\cal U}^1_{j,{\hat n}},{\cal U}^2_{j,{\hat n}}]
+[{\cal U}^1_{j-{\hat n},{\hat n}},{\cal U}^2_{j-{\hat n},{\hat n}}]\right).$$
Note that the initial conditions as described satisfy the Gauss constraints
\footnote{The lattice initial conditions are not completely specified by
requiring that they reduce to the corresponding conditions in the continuum upon
taking the formal continuum limit. This arbitrariness can be removed by a
consistent derivation of
the initial conditions entirely within the lattice theory. Such derivation will 
be presented elsewhere \cite{AlexRaju}.}.

Finally, we comment on numerical integration of the lattice equations of motion.
A method of choice for energy-conserving Hamiltonian equations is the leapfrog 
algorithm,
wherein, the fields are updated at every step of the integration while their
conjugate momenta are kept fixed and vice versa. In our case, the Hamiltonian
depends explicitly on time and the energy is not conserved. 
Nevertheless, only a minor modification of the algorithm is necessary in order 
to maintain the same time step accuracy as in the energy-conserving case.
As explained in \cite{thalgs}, the leapfrog algorithm has the useful
property of respecting Gauss' laws {\it exactly} (regardless of the time step)
for any Hamiltonian which is a sum of potential and kinetic terms.

\section{Outlook}

We have outlined above a procedure to solve for the gauge field configurations 
produced in the collision of the Weizs\"acker--Williams fields of two 
nuclei. These can then be used to compute a large variety of observables 
as a function of proper time $\tau$. An observable that can be computed
directly in our Hamiltonian approach is the energy density 
(and correlations in the energy density). Also, by looking at the 
Fourier decomposition of the co--ordinate space correlators of the electric 
fields, one can determine the energy and number distributions of the 
modes as well as the energy dispersion relation for the soft modes. These 
can be directly related to the multiplicity and transverse energy of 
mini--jets (for a review see Ref.~\cite{Kari}). Some care however must
be exercised in ensuring that the residual gauge freedom of the fields is
fixed properly.

The field configurations generated immediately after the collision are 
completely out of equilibrium. It is an interesting problem to study 
if and how the system approaches equilibrium. This can be done by studying
whether the multiplicity saturates as a function of time. Alternatively,
one can follow the example of the Duke group and study the behaviour of
the Lyapunov exponents as a function of time~\cite{Muller,MullRev}. 
If the system does indeed thermalize, energy density/temperature and 
velocity profiles can be extracted for use as initial conditions in 
hydrodynamic simulations.

A useful check of the results of our proposed simulation (besides the usual 
technical ones) is to reproduce the results of Ref.~\cite{KLW,gyulassy,
DirkYuri} in the ``Abelian limit'' of large transverse momenta, $k_t>>
\alpha_S \mu$. The sensitivity of various observables to the lattice 
spacing and lattice size must also be carefully studied.

There are several open questions which have still to be resolved. Primarily, 
our simulation is completely classical--how accurate is the
classical description? It would be interesting to study whether the Wilson 
renormalization group evolution for the scale $\chi(\eta,Q^2)$ can be 
implemented on the lattice. Another issue is our assumption of boost 
invariance--relaxing this assumption however is not conceptually difficult 
but may be numerically time consuming. 

A more detailed discussion of the above and detailed numerical simulations 
will be presented at a later date~\cite{AlexRaju}.

\section*{Acknowledgments}

One of us (R.V.) would like to thank Miklos Gyulassy, Larry McLerran and 
Berndt M\"uller for useful discussions. He would also
like to thank the organizers of the ``International conference on the
physics and astrophysics of the quark gluon plasma'' held in Jaipur, India, 
from March 17th--21st.


\begin{thebibliography} {99}

\bibitem{QM96}see for instance, the proceedings of {\em Quark Matter 96},
{\em Nucl. Phys.} {\bf A610} (1996).

\bibitem{Wang}X.-N. Wang, {\em Phys. Rep.} {\bf 280} 287 (1997). 

\bibitem{Geiger}K. Geiger, {\em Phys.Rep.} {\bf 258} 237 (1995).

\bibitem{comment}R. Venugopalan, hep-ph/9604209, to appear in {\em 
Comments in Nucl. and Part. Physics}. 

\bibitem{RajLar}
L. McLerran and  R. Venugopalan, {\em Phys. Rev.} {\bf D49} 2233 (1994); 
{\bf D49} 3352 (1994).

\bibitem{AlexLar}J. Jalilian--Marian, A. Kovner, L. McLerran and H. Weigert, 
{\em Phys. Rev.} {\bf D55} 5414 (1997).

\bibitem{Kovchegov}Yu. V. Kovchegov, {\it Phys. Rev.} {\bf D54} 5463 (1996); 
{\bf D55} 5445 (1997).

\bibitem{KLW}A. Kovner, L. McLerran and H. Weigert, {\em Phys. Rev} 
{\bf D52} 3809 (1995); {\bf D52} 6231 (1995).

\bibitem{gyulassy} M. Gyulassy and L. McLerran, nucl-th/9704034.

\bibitem{DirkYuri}Y. V. Kovchegov and D. H. Rischke, hep-ph/9704201.

\bibitem{SerBerDir}S. G. Matinyan, B. M\"uller and D. H. Rischke, nucl-th/
9705024.

\bibitem{Krasnitz}J. Ambjorn and A. Krasnitz, {\em Phys. Lett.} 
{\bf B362} 97 (1995); hep-ph/9705380.

\bibitem{Moore}G. D. Moore, {\em Nucl. Phys.} {\bf B480} 657 (1996).

\bibitem{BFDGL}E.A. Kuraev, L.N. Lipatov and Y.S. Fadin, {\it Zh. Eksp. Teor. 
Fiz}
{\bf 72}, 3 (1977) ({\it Sov. Phys. JETP }{\bf 45}, 1 (1977) ); I.A.
Balitsky and L.N. Lipatov, {\it Sov. J. Nucl. Phys. }{\bf 28} 822 (1978); 
G. Altarelli and G. Parisi, {\it Nucl. Phys.} {\bf B126} 298 (1977); 
Yu.L. Dokshitser,  {\it Sov.Phys.JETP} {\bf 46} 641 (1977). 

\bibitem{Kovner}J. Jalilian--Marian, A. Kovner, A. Leonidov and H. Weigert, 
hep-ph/9701284.

\bibitem{Bj}J. D. Bjorken, {\em Phys.Rev.} {\bf D27} 140 (1983). 

\bibitem{josef}J. Sollfrank, P. Houvinen, M. Kataja, P. V. Ruuskanen, 
M. Prakash and R. Venugopalan, {\em Phys.Rev.} {\bf C55} 392 (1997). 

\bibitem{Sasha}A. Makhlin, hep-ph/9608261.

\bibitem{KS}J. Kogut and L. Susskind, {\em Phys. Rev.} {\bf D11} 395 (1975).

\bibitem{Rajs}R. V. Gavai and R. Venugopalan, {\em Phys. Rev.} {\bf D54} 
5795 (1996).

\bibitem{RajLar2}L. McLerran and R. Venugopalan, nucl-th/9705055, submitted 
to {\em Phys. Lett.} {\bf B}.

\bibitem{thalgs}A. Krasnitz, {\em Nucl. Phys.} {\bf B455} 320 (1995).

\bibitem{Kari}K. E. Eskola, nucl-th/9705027, submitted to {\em Comments 
in Nucl. and Part. Phys.}

\bibitem{Muller}U. Heinz, C.R. Hu, S. Leupold, S.G. Matinian and B. M\"uller,
{\em Phys. Rev.} {\bf D55} 2464 (1997).

\bibitem{MullRev}T. S. Biro, C. Gong, B. M\"uller and A. Trayanov, 
{\em Int. J. Mod. Phys.} {\bf C5} 113 (1994).

\bibitem{AlexRaju}A. Krasnitz and R. Venugopalan, in progress.

\end{thebibliography}
\end{document}